\documentclass[journal]{IEEEtran}
\usepackage{amsfonts}

\newtheorem{theorem}{Theorem}

\newtheorem{remark}[theorem]{Remark}

\input{tcilatex}
\begin{document}

\title{Information Theory and Point Processes}
\author{Ronald Mahler, Random Sets LLC, Eagan, MN, U.S.A.}
\maketitle

\begin{abstract}
This paper addresses theoretically correct vs. incorrect ways to apply
information theory to point processes.
\end{abstract}

\section{Introduction}

\setcounter{page}{1}Point proceess (p.p.) theory addresses the statistical
behavior of randomly varying finite ensembles of points in some state space 
\cite{Daley-Jones}. \ There are multiple formulations of the theory; here
the focus will be on the vector-based version. \ The purpose of the paper is
to address theoretically correct vs. incorrect ways of devising
information-theoretic functionals for vector-based p.p.'s. \ 

Several such functionals were recently introduced in \cite{ClarkEntropy2020}%
. \ In what follows it will be demonstrated that the theoretical approach in
that paper is incorrect and that, as a consequence, Eqs. (42-54) of \cite%
{ClarkEntropy2020} are mathematically undefined when employed in typical
engineering applications. \ In particular, in these applications the key
Eqs. (42,47,51) involve summations of terms with incommensurable units of
measurement; and this remains the case even if the theoretically correct
approach is applied instead. \ 

The paper is organized as follows: \ summary of the approach in \cite%
{ClarkEntropy2020} (Section \ref{A-Theory}); description of the correct
approach (Section \ref{A-PPInfo}); and critique of the mathematically
undefined formulas (Section \ref{A-Undefined}).\ \ \ 

\section{Background \label{A-Theory}}

\subsection{Vector-Based Point Processes \label{A-Theory-AA-PP}}

Let \ $\mathfrak{X}$ \ be a topological space\ (hereafter referred to as the
\textquotedblleft base space\textquotedblright ),\ endowed with a measure \ $%
\lambda _{\mathfrak{X}}(B)$ \ (hereafter referred to as the
\textquotedblleft base measure\textquotedblright ) defined on the
Borel-measurable subsets \ $B$ \ of \ $\mathfrak{X}$. \ In typical
engineering applications \ $\mathfrak{X}$ \ is a region of a Euclidean space
with unit of measurement \ $\iota $, which is what will be assumed
hereafter. \ In this case \ $\lambda _{\mathfrak{X}}$ \ is Lebesgue measure
and the unit of measurement of \ $\lambda _{\mathfrak{X}}(B)$ \ is \ $\iota $%
. \ \ \ 

According to \cite{ClarkEntropy2020}, p. 1157, a p.p. is a random variable
(r.v.) \ $\Phi $ \ on the disjoint-union space \ $\mathfrak{X}^{\infty
}=\biguplus_{n\geq 0}\mathfrak{X}^{n}$ \ of finite ordered sequences \ $%
\varphi =(x_{1},...,x_{n})$ \ with \ $x_{1},...,x_{n}\in \mathfrak{X}$; and
where for \ $n=0$, \ $\varphi =\phi $ \ denotes the empty sequence; and
where we define \ $|\varphi |=n$. \ Thus $\mathfrak{X}^{\infty }$ \ is the
topological sum of \ $\mathfrak{X}^{0}=\{\phi \},\mathfrak{X},\mathfrak{X}%
^{2},...$; the open subsets \ $\mathfrak{O}$ \ of which are \ $\mathfrak{O}%
=\biguplus_{n\geq 0}(\mathfrak{O}\cap \mathfrak{X}^{n}\mathcal{)}$ \ such
that \ $\mathfrak{O}\cap \mathfrak{X}^{n}$ \ are open in \ $\mathfrak{X}^{n}$%
; and thus where the \ $\mathfrak{O}\cap \mathfrak{X}^{n}$ \ are generated
by the Cartesian products \ $O_{1}\times ...\times O_{n}$ \ for open \ $%
O_{1},...,O_{n}\subseteq \mathfrak{X}$. \ The \ $\mathfrak{X}^{n}$ \ for \ $%
n\geq 0$ \ are both open and closed in \ $\mathfrak{X}^{\infty }$, so that \ 
$\mathfrak{X}^{\infty }$ \ is topologically disconnected. \ 

The statistics of \ $\Phi $ \ are described by its probability measure \ $%
P_{\Phi }(\mathfrak{B})$ \ where \ $\mathfrak{B}$ \ is a Borel-measurable
subset of \ $\mathfrak{X}^{\infty }$. \ That is, $\mathfrak{B}%
=\biguplus_{n\geq 0}(\mathfrak{B}\cap \mathfrak{X}^{n}\mathcal{)}$ \ such
that \ $\mathfrak{B}\cap \mathfrak{X}^{n}$ \ are measurable in \ $\mathfrak{X%
}^{n}$; and thus where the \ $\mathfrak{B}\cap \mathfrak{X}^{n}$ \ are
generated by the \ $B_{1}\times ...\times B_{n}$ \ for measurable \ $%
B_{1},...,B_{n}\subseteq \mathfrak{X}$. \ The measure \ $P_{\Phi }$ \ is
assumed to be symmetric---i.e., if \ $\mathfrak{B}^{\prime }$ \ is
constructed from \ $\mathfrak{B}$ \ by permuting some or all of the entries
in some or all of the \ $\varphi \in \mathfrak{B}$, \ then \ $P_{\Phi }(%
\mathfrak{B}^{\prime })=P_{\Phi }(\mathfrak{B})$. \ The projection measure \ 
\begin{equation}
P_{\Phi }^{(n)}(\mathfrak{B})=P_{\Phi }(\mathfrak{B}\cap \mathfrak{X}^{n})
\label{eq-ProjMeas}
\end{equation}%
describes the realizations of \ $\Phi $ \ with \ $n$ \ elements;\ and \ $%
P_{\Phi }(\mathfrak{B})=\sum_{n\geq 0}P_{\Phi }^{(n)}(\mathfrak{B})$.

\subsection{Probability Generating Functional of a p.p. \label%
{A-Theory-AA-PGFL}}

According to \cite{Moyal},\ p. 2, the probability generating functional
(p.g.f{}l.) \ $\mathcal{G}_{\Phi }(h)$ \ of a p.p. \ $\Phi $ \ was
introduced by Bartlett and Kendall in the late 1940s and is (\cite{Moyal}; 
\cite{Daley-Jones}; \cite{ClarkEntropy2020}, Eq. (29)): 
\begin{equation}
\mathcal{G}_{\Phi }(h)=\sum_{n\geq 0}\int \left(
\prod_{i=1}^{n}h(x_{i})\right) P_{\Phi }^{(n)}(dx_{1},...,dx_{n})
\label{eq-PGFL}
\end{equation}%
where the \textquotedblleft test function\textquotedblright\ \ $h:\mathfrak{X%
}\rightarrow \lbrack 0,1]$ \ is unitless. \ Note that \ $0\leq \mathcal{G}%
_{\Phi }(h)\leq \mathcal{G}_{\Phi }(1)=P_{\Phi }(\mathfrak{X}^{\infty })=1$.

\subsection{Chain Differential of a p.g.f{}l.\ \label{A-Theory-AA-Chain}}

If \ $F(h)\geq 0$ \ is a functional on \ $h:\mathfrak{X}\rightarrow \lbrack
0,1]$, then its chain differential is (\cite{Bernard}; \cite%
{ClarkEntropy2020}, Eq. (35)):%
\begin{equation}
\delta F(h;\eta )=\lim_{i\rightarrow \infty }\frac{F(h+\varepsilon _{i}\eta
_{i})-F\left( h\right) }{\varepsilon _{i}}  \label{eq-ChainDiff}
\end{equation}%
if the limits exist and are identical for any sequences \ $\varepsilon
_{i}\rightarrow 0$ \ and \ $\eta _{i}\rightarrow \eta $ \ (pointwise).%
\footnote{%
The chain differential of \ $F$ \ is a modification of the G\^{a}teaux
differential of $F$, which is defined as \ $\lim_{\varepsilon \rightarrow
0^{+}}\varepsilon ^{-1}(F(h+\varepsilon \eta )-F\left( h\right) )$.} \ Also,
\ $\delta ^{n}F(h;\eta _{1},...,\eta _{n})$ \ is defined recursively by \ $%
(\delta ^{n}F)(h;\eta _{1},...,\eta _{n})=\delta F_{n-1}(h;\eta _{n})$ \
where \ $F_{n-1}(h)=\delta ^{n-1}F(h;\eta _{1},...,\eta _{n-1})$. \ By
convention, \ $\delta ^{0}F(h)=F(h)$. \ 

Assume that \ $F(h)$ \ is well-behaved enough that: \ 

(a) $\eta \mapsto \delta F(h;\eta )$ \ is linear and continuous,\footnote{%
Because of these assumptions \ $\delta F(h;\eta )$ \ is a chain derivative 
\cite{Bernard}, not just a chain differential.} in which case \ $\mu
_{F,h}:B\mapsto \delta F(h;\mathbf{1}_{B})$ \ is a measure on \ $\mathfrak{X}
$; and

(b) \ $\mu _{F,h}$ \ is absolutely continuous w/r/t (with respect to) \ $%
\lambda _{\mathfrak{X}}$. \ 

Then \ $\delta F(h;\delta _{x})$ \ is shorthand for the Radon-Nikod\'{y}m
derivative \ $(d\mu _{F,h}/d\lambda _{\mathfrak{X}})(x)$ \ (\cite%
{ClarkEntropy2020}, Eq. (39)), where \ $\delta _{x}(y)=(d\Delta
_{x}/d\lambda _{\mathfrak{X}})(y)$ \ is the Dirac delta function on \ $%
\mathfrak{X}$, $\ \Delta _{x}(B)=\mathbf{1}_{B}(x)$ \ is the Dirac measure,
and \ $\mathbf{1}_{B}$\ \ is the set indicator function of \ $B$.

$P_{\Phi }$ \ and \ $\mathcal{G}_{\Phi }$ \ are related by (\cite{Moyal},
Eq. (4.14); \cite{ClarkEntropy2020}, Eq. (36)):\ 
\begin{equation}
P_{\Phi }^{(n)}(B_{1}\times ...\times B_{n})=\frac{1}{n!}\delta ^{n}\mathcal{%
G}_{\Phi }(0;\mathbf{1}_{B_{1}},...,\mathbf{1}_{B_{n}})  \label{eq-Moyal}
\end{equation}%
for measurable \ $B_{1},...,B_{n}\subseteq \mathfrak{X}$, \ with \ $P_{\Phi
}^{(0)}(\{\phi \})=\mathcal{G}_{\Phi }(0)$. \ For \ $n\geq 0$ \ define (\cite%
{ClarkEntropy2020}, Eq. (39)): 
\begin{equation}
p_{\Phi }^{(n)}(x_{1},...,x_{n})=\frac{1}{n!}\delta ^{n}\mathcal{G}_{\Phi
}(0;\delta _{x_{1}},...,\delta _{x_{n}}),  \label{eq-Clark}
\end{equation}%
so that \ $p_{\Phi }^{(0)}=\mathcal{G}_{\Phi }(0)$. \ The left side of (\ref%
{eq-Clark}) is the family of Janossy densities of \ $\Phi $ \ \cite%
{Daley-Jones} indexed by \ $n\geq 1$. \ Because \ $P_{\Phi }$ \ is
symmetric, \ $p_{\Phi }^{(n)}(x_{1},...,x_{n})$ \ is symmetric w/r/t \ \ $%
x_{1},...,x_{n}$ \ for each \ $n\geq 2$.

The \ $p_{\Phi }^{(n)}(x_{1},...,x_{n})$ \ exist only if \ $\Phi $ \ is
\textquotedblleft simple\textquotedblright ---i.e., if \ \ $x_{1},...,x_{n}$
\ are distinct\ for any realizations \ $\Phi =(x_{1},...,x_{n})$ \ with \ $%
n\geq 2$ \ \cite{Daley-Jones}.\ Thus all p.p.'s in \cite{ClarkEntropy2020}
have implicitly been assumed to be simple.

The chain differential, G\^{a}teaux differential, and Frech\'{e}t derivative
of a p.g.f{}l. \ $\mathcal{G}_{\Phi }$ \ exist and are equal; and are
equivalent to the Volterra functional derivative\footnote{%
Functional derivative: see \cite{Mah-Artech}, p. 375.} $\ (\delta \mathcal{G}%
_{\Phi }/\delta x)(h)$ \ of \ $\mathcal{G}_{\Phi }$ in the sense that%
\begin{equation}
\delta \mathcal{G}_{\Phi }(h;\eta )=\int \eta (x)\cdot \frac{\delta \mathcal{%
G}_{\Phi }}{\delta x}(h)dx
\end{equation}%
and thus that\ \ $\delta \mathcal{G}_{\Phi }(h;\delta _{x})=(\delta \mathcal{%
G}_{\Phi }/\delta x)(h)$.\footnote{%
This all follows from the fact that a p.g.f{}l. is a functional\ power
series and thus is, in this sense, analytic \cite{MahSensors2019}.} \ It is
well known that the Frech\'{e}t derivative admits a chain rule. \ The chain
differential was introduced in 2005 in \cite{Bernard} to permit a chain rule
for functions that are not Frech\'{e}t differentiable. \ The need for it in
a p.g.f{}l. context is therefore unclear.

For a more detailed critique of the vector-based p.p. framework, see \cite%
{MahSensors2019}.\ 

\section{Information Theory and p.p.'s \label{A-PPInfo}}

The purpose of this section is to describe the theoretically correct way to
apply information theory to p.p.'s. \ Given the mathematical formulation in 
\cite{ClarkEntropy2020}, \textquotedblleft theoretically
correct\textquotedblright\ means \textquotedblleft correctly
measure-theoretic.\textquotedblright

\begin{remark}
\label{Rem-Unnecessary}This section is not needed to understand why Eqs.
(42-54) of \cite{ClarkEntropy2020} are erroneous---see (\ref{eq-42},\ref%
{eq-Laplace},\ref{eq-cumulant}). \ Rather, it is required for the
demonstration in Remark \ref{Rem-Substitution} that these errors cannot be
corrected by employing the theoretically correct approach described in this
section.\ 
\end{remark}

It is organized as follows: \ Lesbegue integration (Section \ref%
{A-PPInfo-AA-Lebesgue}); differential entropy as a simple example (Section %
\ref{A-PPInfo-AA-InfoTheory}); measure and integration for p.p.'s (Section %
\ref{A-PPInfo-AA-Measure}); the probability density function of a p.p.
(Section \ref{A-PPInfo-AA-RadNik}); and optimal state estimation for p.p.'s
(Section \ref{A-PPInfo-AA-E}).

\subsection{Lebesgue Integration \label{A-PPInfo-AA-Lebesgue}}

Let \ $\mu (B)$ \ be a unitless measure of the Lebesgue-measurable subsets \ 
$B\subseteq \mathfrak{X}$ \ and assume that it is absolutely continuous
w/r/t \ $\lambda _{\mathfrak{X}}$---i.e., \ $\lambda _{\mathfrak{X}}(B)=0$ \
implies \ $\mu (B)=0$.\ \ Then by the Radon-Nikod\'{y}m theorem, there is an
almost everywhere unique Lebesgue-integrable function \ $d\mu /d\lambda _{%
\mathfrak{X}}:\mathfrak{X}\rightarrow \mathbb{R}^{+}$---the Radon-Nikod\'{y}%
m derivative of \ $\mu $ \ w/r/t the base measure \ $\lambda _{\mathfrak{X}}$%
---such that \ 
\begin{equation}
\mu (B)=\int_{B}\frac{d\mu }{d\lambda _{\mathfrak{X}}}(x)dx
\end{equation}%
for all \ $B$ \ and where \ $\lambda _{\mathfrak{X}}(dx)$ \ has been
abbreviated as \ $dx$. \ Because \ $\mu (B)$ \ is unitless and \ $dx$ \ has
unit \ $\iota $, \ $(d\mu /d\lambda _{\mathfrak{X}})(x)$ \ must have unit \ $%
\iota ^{-1}$.

\begin{remark}
\label{Rem-Units}Since \ $\mathcal{G}_{\Phi }$ \ and thus \ $\mu _{\mathcal{G%
}_{\Phi },h}$ \ are unitless, it follows that \ $\delta \mathcal{G}_{\Phi
}(h;\delta _{x})=(d\mu _{\mathcal{G}_{\Phi },h}/d\lambda _{\mathfrak{X}})(x)$%
\ \ has unit \ $\iota ^{-1}$ \ and \ $\delta ^{n}\mathcal{G}_{\Phi
}(h;\delta _{x_{1}},...,\delta _{x_{n}})$\ \ has unit \ $\iota ^{-n}$.
\end{remark}

\begin{remark}
\label{Rem-Unit}From this it follows that $\ p_{\Phi
}^{(n)}(x_{1},...,x_{n}) $ \ has unit \ $\iota ^{-n}$ \ for \ $n\geq 0$.\ \ 
\end{remark}

\subsection{Simple Example: \ Differential Entropy \label%
{A-PPInfo-AA-InfoTheory}}

Let $\ \mathbf{X}\in \mathfrak{X}$ \ be an r.v. with probability measure \ $%
P_{\mathbf{X}}(B)$ \ and probability density function (p.d.f.) \ $f_{\mathbf{%
X}}(x)=(dP_{\mathbf{X}}/d\lambda _{\mathfrak{X}})(x)$,\ where \ $dP_{\mathbf{%
X}}/d\lambda _{\mathfrak{X}}$ \ is the Radon-Nikod\'{y}m derivative of \ $P_{%
\mathbf{X}}$ \ w/r/t $\ \lambda _{\mathfrak{X}}$. \ Then the \textit{%
differential entropy} (DE) of \ $\mathbf{X}$ \ is:%
\begin{equation}
DE(f_{\mathbf{X}})=-\int f_{\mathbf{X}}(x)\log f_{\mathbf{X}}(x)dx=-\int
\log f_{\mathbf{X}}(x)P_{\mathbf{X}}(dx)\mathbf{.}
\end{equation}%
The DE has two well-known limitations as a quantifier of information. \
First, it is not invariant w/r/t change of coordinates. \ Second and in
particular, it is\ undefined if $\ \mathfrak{X}$ \ and thus \ $f_{\mathbf{X}%
} $ \ have units of measurement and thus $\log f_{\mathbf{X}}(\mathbf{x})$ \
is undefined. \ A minimum requirement for any quantifier of information (or
entropy) should be that its numerical value does not change if (for example)
one converts from metric to English units. \ 

Csisz\'{a}r information functionals, such as the Kullback-Leibler divergence%
\begin{equation}
KL(f_{\mathbf{X}};f_{0})=\int \log \left( \frac{f_{\mathbf{X}}(x)}{f_{0}(x)}%
\right) P_{\mathbf{X}}(dx),
\end{equation}%
do not have these limitations---see, e.g., Eq. (3) of \cite{QianTSP-2010}.

Nevertheless, the DE provides a simple illustration of how to correctly
apply information theory to p.p.'s.\ \ Here we employ it to summarize this
approach, assuming the measure-theoretic results to be established in
Sections \ref{A-PPInfo-AA-Measure} and \ref{A-PPInfo-AA-RadNik}.

The DE of a p.p. \ $\Phi $ \ must be%
\begin{equation}
DE(f_{\Phi })=-\int f_{\Phi }(\varphi )\log f_{\Phi }(\varphi )d\varphi
=-\int \log f_{\Phi }(\varphi )P_{\Phi }(d\varphi )  \label{eq-DE-PP}
\end{equation}%
where $\ f_{\Phi }(\varphi )=(dP_{\Phi }/d\lambda _{\mathfrak{X}^{\infty
}})(\varphi )$ \ and where \ $\lambda _{\mathfrak{X}^{\infty }}(\mathfrak{B}%
) $\ \ is a measure \textit{on the p.p. state space }\ $\mathfrak{X}^{\infty
}$. \ But (\ref{eq-DE-PP}) is meaningless unless we answer the following
question: \ \textit{What is}\ $\ \lambda _{\mathfrak{X}^{\infty }}$\textit{?}
\ 

To be conceptually consistent, \ \ $\lambda _{\mathfrak{X}^{\infty }}$ \
must be an extension of \ $\lambda _{\mathfrak{X}}$ \ to \ $\mathfrak{X}%
^{\infty }$. \ If \ $\mathfrak{X}$\ \ has unit of measurement \ $\iota $ \
then the simplest extension is \ $\lambda _{\mathfrak{X}^{\infty }}=\lambda
_{c}$\ \ with \ $c>0$ \ as given below in (\ref{eq-Induced}). \ This leads,
in (\ref{eq-RadNik}), to the formula \ $f_{\Phi }(\varphi )=c^{|\varphi
|}p_{\Phi }^{(|\varphi |)}(\varphi )$---which in turn, means that if \ $%
\mathfrak{X}$ \ has a unit then \ $p_{\Phi }^{(|\varphi |)}(\varphi )$ \ is
not the p.d.f. of \ $\Phi $. \ That is: \ 

\begin{remark}
\label{Rem-MainPoint}Any p.p. quantifier of information (or entropy) that is
measure-theoretic and generally applicable must be defined using\textit{\ \ }%
$f_{\Phi }(\varphi )$\textit{\ \ }rather than \textit{\ }$p_{\Phi
}^{(|\varphi |)}(\varphi )$\textit{.}
\end{remark}

\subsection{Measure and Integration for p.p.'s\ \label{A-PPInfo-AA-Measure}}

Just as the $\sigma $-algebra of \ $\mathfrak{X}^{\infty }$ \ is an
extension of the $\sigma $-algebra of \ $\mathfrak{X}$, so \ $\lambda _{%
\mathfrak{X}^{\infty }}$ \ is the following well-known extension of \ $%
\lambda _{\mathfrak{X}}$ \ to $\mathfrak{X}^{\infty }$ (\cite{Mah-Artech},
p. 715): \ 
\begin{equation}
\lambda _{c}(\mathfrak{B})=\mathbf{1}_{\mathfrak{B}}(\phi )+\sum_{n\geq 1}%
\frac{\lambda _{\mathfrak{X}}^{n}(\mathfrak{B}\cap \mathfrak{X}^{n})}{c^{n}}.
\label{eq-Induced}
\end{equation}%
Here, \ $c>0$ \ has unit \ $\iota $; $\ \lambda _{\mathfrak{X}}^{1}=\lambda
_{\mathfrak{X}}$; \ and for \ $n\geq 2$, \ $\lambda _{\mathfrak{X}}^{n}$ \
is the extension of \ $\lambda _{\mathfrak{X}}$ \ to \ $\mathfrak{X}^{n}$,
in which case \ $\lambda _{\mathfrak{X}}^{n}(B_{1}\times ...\times B_{n})$ \
has unit \ $\iota ^{n}$. \ Because of \ \ $c$, \ the summation in (\ref%
{eq-Induced}) is mathematically well-defined since its terms are unitless. \
Like \ $\lambda _{\mathfrak{X}}$, \ $\lambda _{c}$ \ is a possibly
infinite-valued measure. \ Unlike \ $\lambda _{\mathfrak{X}}$, it is
unitless. \ 

The projection measures of \ $\lambda _{c}$ \ are \ $\lambda _{c}^{(0)}(%
\mathfrak{B})=\mathbf{1}_{\mathfrak{B}}(\phi )$ \ and \ $\lambda _{c}^{(n)}(%
\mathfrak{B})=\lambda _{\mathfrak{X}}^{n}(\mathfrak{B}\cap \mathfrak{X}%
^{n})/c^{n}$ \ for \ $n\geq 1$. \ In particular, \ $\lambda _{c}^{(1)}(%
\mathfrak{B})=\lambda _{\mathfrak{X}}(\mathfrak{B}\cap \mathfrak{X})/c$. \ 

Let \ $f(\varphi )$ \ be a nonnegative unitless function of \ $\varphi $. \
Then its integral w/r/t \ $\lambda _{c}$ \ within \ $\mathfrak{B}\subseteq 
\mathfrak{X}^{\infty }$ \ is%
\begin{eqnarray}
&&\int_{\mathfrak{B}}f(\varphi )\lambda _{c}(d\varphi )  \label{eq-Integral0}
\\
&=&f(\phi )\cdot \mathbf{1}_{\mathfrak{B}}(\phi )+\sum_{n\geq 1}\frac{1}{%
c^{n}}\int_{\mathfrak{B}\cap \mathfrak{X}^{n}}f(x_{1},...,x_{n})dx_{1}\cdots
dx_{n}  \nonumber
\end{eqnarray}%
where \ $\lambda _{\mathfrak{X}}^{n}(dx_{1},...,dx_{n})$ \ has been
abbreviated as \ $dx_{1}\cdots dx_{n}$. \ Thus \ $\lambda _{c}(\mathfrak{B}%
)=\int_{\mathfrak{B}}\lambda _{c}(d\varphi )$.\ \ 

\subsection{Probability Density Function (p.d.f.) of a p.p. \label%
{A-PPInfo-AA-RadNik}}

If \ $P_{\Phi }$ \ is absolutely continuous w/r/t \ $\lambda _{c}$ \ then
its p.d.f.\ is \ $f_{\Phi }(\varphi )=dP_{\Phi }/d\lambda _{c})(\varphi )$,
which is characterized by the Radon-Nikod\'{y}m theorem%
\begin{eqnarray}
P_{\Phi }(\mathfrak{B}) &=&\int_{\mathfrak{B}}f_{\Phi }(\varphi )\lambda
_{c}(d\varphi )=\int_{\mathfrak{B}}\frac{dP_{\Phi }}{d\lambda _{c}}(\varphi
)\lambda _{c}(d\varphi ) \\
&=&\frac{dP_{\Phi }}{d\lambda _{c}}(\phi )\cdot \mathbf{1}_{\mathfrak{B}%
}(\phi )  \label{eq-ProbMeas} \\
&&+\sum_{n\geq 1}\frac{1}{c^{n}}\int_{\mathfrak{B}\cap \mathfrak{X}^{n}}%
\frac{dP_{\Phi }}{d\lambda _{c}}(x_{1},...,x_{n})dx_{1}\cdots dx_{n} 
\nonumber
\end{eqnarray}%
where the restiction of\ \ $dP_{\Phi }/d\lambda _{c}$ \ to \ $\mathfrak{X}%
^{n}$ \ is equal to \ $dP_{\Phi }^{(n)}/d\lambda _{c}$. \ By (\ref%
{eq-ProbMeas}), (\ref{eq-ProjMeas}), and (\ref{eq-Moyal}), for \ $n\geq 1$ \
the projection measures \ $P_{\Phi }^{(n)}$ \ of \ $P_{\Phi }$ \ are given
by 
\begin{eqnarray}
&&\frac{1}{n!}\delta ^{n}\mathcal{G}_{\Phi }(0;\mathbf{1}_{B_{1}},...,%
\mathbf{1}_{B_{n}}) \\
&=&\frac{1}{c^{n}}\int_{B_{1}\times ...\times B_{n}}\frac{dP_{\Phi }^{(n)}}{%
d\lambda _{c}}(x_{1},...,x_{n})dx_{1}\cdots dx_{n}  \nonumber \\
&=&\frac{1}{c^{n}}\int_{B_{1}\times ...\times B_{n}}f_{\Phi
}(x_{1},...,x_{n})dx_{1}\cdots dx_{n}.
\end{eqnarray}%
From this and (\ref{eq-Clark}) it follows that%
\begin{equation}
p_{\Phi }^{(n)}(x_{1},...,x_{n})=\frac{1}{c^{n}}\cdot f_{\Phi
}(x_{1},...,x_{n})
\end{equation}%
and thus that the p.d.f. of \ $P_{\Phi }$ \ is the unitless function \ 
\begin{equation}
f_{\Phi }(\varphi )=c^{|\varphi |}p_{\Phi }^{(|\varphi |)}(\varphi ).
\label{eq-RadNik}
\end{equation}

\begin{remark}
\label{Rem-Error}Since \ $c$ \ has unit \ $\iota $, \ \ $f_{\Phi }(\varphi
)\neq p_{\Phi }^{(|\varphi |)}(\varphi )$ \ even if \ $c=1\cdot \iota $ \
since \ $f_{\Phi }(\varphi )$ \ is unitless and \ $p_{\Phi }^{(|\varphi
|)}(\varphi )$ \ is not. \ Thus the family \ $p_{\Phi
}^{(n)}(x_{1},...,x_{n})$ \ of Janossy densities on \ $\mathfrak{X}^{n}$ \
for \ $n\geq 0$ \ is not the same thing as the p.d.f. \ $f_{\Phi }(\varphi )$
\ of \ $\Phi $, which is a single density\ on \ $\mathfrak{X}^{\infty }$. \ 
\end{remark}

The application of advanced Gibbs statistical sampling techniques to exact
closed-form approximations of \ $f_{\Phi }(\varphi )$---or more precisely,
of \ $f_{\Phi }(X)$ \ as defined in Section \ref{A-PPInfo-AA-E}---has led to
implementations of the generalized labeled multi-Bernoulli (GLMB)\ filter
that are capable of simultaneous real-time tracking of over a million
targets in significant clutter using off-the-shelf computing equipment \cite%
{Beard2020}. \ 

\begin{remark}
\label{Rem-Equiv}It follows that (\ref{eq-ProbMeas}) can be rewritten as%
\begin{equation}
P_{\Phi }(\mathfrak{B})=\sum_{n\geq 0}\int_{\mathfrak{B}\cap \mathfrak{X}%
^{n}}p_{\Phi }^{(n)}(x_{1},...,x_{n})dx_{1}\cdots dx_{n}.
\end{equation}
\end{remark}

\begin{remark}
\label{Rem-Integral}It also follows that if \ $n\geq 1$ then%
\begin{equation}
P_{\Phi }^{(n)}(dx_{1},...,dx_{n})=p_{\Phi
}^{(n)}(x_{1},...,x_{n})dx_{1}\cdots dx_{n}.  \label{eq-Integral}
\end{equation}
\end{remark}

For, from (\ref{eq-Integral0}) and (\ref{eq-RadNik}) we have%
\begin{eqnarray}
&&\int f(x_{1},...,x_{n})P_{\Phi }^{(n)}(dx_{1},...,dx_{n}) \\
&=&\int f(x_{1},...,x_{n})\cdot \frac{dP_{\Phi }^{(n)}}{d\lambda _{c}}%
(x_{1},...,x_{n})\lambda _{c}(dx_{1},...,dx_{n})  \nonumber \\
&=&\frac{1}{c^{n}}\int f(x_{1},...,x_{n})\cdot c^{n}p_{\Phi
}^{(n)}(x_{1},...,x_{n})dx_{1}\cdots dx_{n} \\
&=&\int f(x_{1},...,x_{n})\cdot p_{\Phi }^{(n)}(x_{1},...,x_{n})dx_{1}\cdots
dx_{n}.
\end{eqnarray}

\subsection{Optimal State Estimation for p.p.'s \label{A-PPInfo-AA-E}}

The most probable realization of \ $\Phi $ \ is the maximum a posteriori
(MAP) estimate extracted from (\ref{eq-RadNik}):%
\begin{equation}
(\hat{x}_{1},...,\hat{x}_{\hat{n}})_{c}=\arg
\sup_{n,x_{1},...,x_{n}}c^{n}p_{\Phi }^{(n)}(x_{1},...,x_{n}).
\label{eq-MAP}
\end{equation}%
Since each \ $c>0$ \ determines a different most-probable estimate, (\ref%
{eq-MAP}) is essentially useless unless we answer the following question: \ 
\textit{What is the best choice for} \ $c$\textit{?}

To answer it, make the following changes from vector to finite-set notation.
\ Assume that \ $x_{1},...,x_{n}$ \ are distinct and let \ $%
X=\{x_{1},...,x_{n}\}$. \ Write \ $f_{\Phi }(X)=n!\cdot p_{\Phi
}^{(n)}(x_{1},...,x_{n})$ \ and \ $|X|=n$, in which case (\ref{eq-MAP})
becomes%
\begin{equation}
\hat{X}_{c}=\arg \sup_{X}\frac{c^{|X|}}{|X|!}f_{\Phi }(X).
\end{equation}%
According to the analysis of \cite{Mah-Artech}, pp. 499-500, for this
estimate to be accurate the magnitude of \ $c$ \ should be approximately
equal to the accuracy with which individual states \ $x\in \mathfrak{X}$ \
are to be estimated. \ 

\begin{remark}
\label{Rem-NotUnity}Thus \ $c\neq 1\cdot \iota $ \ in general. \ 
\end{remark}

\section{Mathematically Undefined Formulas \label{A-Undefined}}

We are now in a position to demonstrate that Eqs. (42-54) of \cite%
{ClarkEntropy2020} are mathematically undefined. \ 

Begin by inspecting the key formula Eq. (42), the \textquotedblleft
Information generating functional for entropy\textquotedblright : \ 
\begin{eqnarray}
\mathcal{G}_{\Phi }^{\alpha }(h) &=&\sum_{n\geq 0}\int \left(
\prod_{i=1}^{n}h(x_{i})\right)  \label{eq-42} \\
&&\cdot p_{\Phi }^{(n)}(x_{1},...,x_{n})^{-\alpha }P_{\Phi
}^{(n)}(dx_{1},...,dx_{n}).  \nonumber
\end{eqnarray}%
Since \ $P_{\Phi }^{(n)}$ \ is a unitless measure and since by Remark \ref%
{Rem-Unit} \ $p_{\Phi }^{(n)}(x_{1},...,x_{n})^{-\alpha }$ \ has unit \ $%
\iota ^{n\alpha }$, then as long as \ $\alpha \neq 0$ the summation is
mathematically undefined since its terms have different units of measurement
for \ $n\geq 0$---and thus are incommensurable. \ 

To see this more explicitly note that, by (\ref{eq-Integral}), we can
rewrite\ (\ref{eq-42}) as:

\begin{eqnarray}
\mathcal{G}_{\Phi }^{\alpha }(h) &=&\sum_{n\geq 0}\int \left(
\prod_{i=1}^{n}h(x_{i})\right)  \label{eq-42b} \\
&&\cdot p_{\Phi }^{(n)}(x_{1},...,x_{n})^{1-\alpha }dx_{1}\cdots dx_{n} 
\nonumber
\end{eqnarray}%
where \ $p_{\Phi }^{(n)}(x_{1},...,x_{n})^{1-\alpha }dx_{1}\cdots dx_{n}$ \
has unit \ $\iota ^{-n(1-\alpha )}\cdot \iota ^{n}=\iota ^{n\alpha }$. \ 

\begin{remark}
\label{Rem-Substitution}This error cannot be corrected by substituting \ $%
f_{\Phi }(x_{1},...,x_{n})$ \ in place of \ $p_{\Phi
}^{(n)}(x_{1},...,x_{n}) $---indeed, it becomes worse. \ For then \ $p_{\Phi
}^{(n)}(x_{1},...,x_{n})^{1-\alpha }dx_{1}\cdots dx_{n}$ \ becomes \ $%
c^{n(1-\alpha )}p_{\Phi }^{(n)}(x_{1},...,x_{n})^{1-\alpha }dx_{1}\cdots
dx_{n}$, \ which has unit \ $\iota ^{n(1-\alpha )}\cdot \iota ^{-n(1-\alpha
)}\cdot \iota ^{n}=\iota ^{n}$ \ regardless of the values of both \ $\alpha $
\ and \ $c$. \ In particular, the error remains even when \ $\alpha =0$. \ 
\end{remark}

\begin{remark}
\label{Rem-InReponse}It might be argued that \ $c=1\cdot \iota $ \ \
suffices as an engineering simplification or approximation, but this is not
the case. \ By Remark \ref{Rem-NotUnity}, \ $c\neq 1\cdot \iota $ \ in
general if we are to find the most probable estimate of \ $\Phi $---i.e.,
find the\ best estimate for engineering purposes.
\end{remark}

These remarks apply with full force to:

\begin{enumerate}
\item Key formula Eq. (47), the \textquotedblleft Laplace information
functional for entropy\textquotedblright :%
\begin{eqnarray}
\mathcal{L}_{\Phi }^{\alpha }(f) &=&\sum_{n\geq 0}\int \exp \left(
-\sum_{i=1}^{n}f(x_{i})\right)  \label{eq-Laplace} \\
&&\cdot p_{\Phi }^{(n)}(x_{1},...,x_{n})^{-\alpha }P_{\Phi
}^{(n)}(dx_{1},...,dx_{n})  \nonumber
\end{eqnarray}%
(for functions \ $f:\mathfrak{X}\rightarrow \mathbb{R}^{+}$),\ which has the
general form \ $\mathcal{L}_{\Phi }^{\alpha }(f)=\mathcal{G}_{\Phi }^{\alpha
}(e^{-f})$ \ where \ $e^{-f}(x)=e^{-f(x)}$. \ 

\item Key formula Eq. (51), the \textquotedblleft cumulant information
functional\textquotedblright :%
\begin{equation}
\mathcal{W}_{\Phi }^{\alpha }(f)=\log \mathcal{L}_{\Phi }^{\alpha }(f).
\label{eq-cumulant}
\end{equation}

\item Any formula defined in terms of \ $\mathcal{G}_{\Phi }^{\alpha }(h)$,
\ $\mathcal{L}_{\Phi }^{\alpha }(h)$, \ or \ $\mathcal{W}_{\Phi }^{\alpha
}(h)$---i.e., Eqs. (42-54).
\end{enumerate}

Additional errors should be noted. \ Consider the \textquotedblleft Shannon
entropy,\textquotedblright\ Eq. (43), which from Eq. (42) can be written as%
\begin{eqnarray}
&&\left[ \frac{\partial }{\partial \alpha }\mathcal{G}_{\Phi }^{\alpha }(1)%
\right] _{\alpha =0}  \label{eq-Shannon} \\
&=&-\sum_{n\geq 0}\int \log p_{\Phi }^{(n)}(x_{1},...,x_{n})P_{\Phi
}^{(n)}(dx_{1},...,dx_{n})  \nonumber \\
&=&-\int \log p_{\Phi }^{(|\varphi |)}(\varphi )P_{\Phi }^{(|\varphi
|)}(d\varphi ).
\end{eqnarray}%
By Remark \ref{Rem-Error}, this is a theorectically erroneous version of the
differential entropy formula (\ref{eq-DE-PP}) since \ $p_{\Phi }^{(|\varphi
|)}(\varphi )\neq f_{\Phi }(\varphi )$. \ It is also mathematically
undefined since, by Remark \ref{Rem-Unit}, \ $p_{\Phi
}^{(n)}(x_{1},...,x_{n})$ \ has unit \ $\iota ^{-n}$ \ and thus \ $\log
p_{\Phi }^{(|\varphi |)}(\varphi )$ \ is mathematically undefined. \ 

The same is true of the \textquotedblleft Shannon entropy
moments,\textquotedblright\ Eq. (48):\footnote{%
As originally written, Eq. (48) had a typo: \ $\mathcal{L}_{\Phi }^{\alpha
}(f)$ \ should have been \ $\mathcal{L}_{\Phi }^{\alpha }(0)$.}%
\begin{eqnarray}
\mathbb{E}_{\Phi }\left[ (\log p_{\Phi })^{m}\right] &=&(-1)^{m}\left[ \frac{%
\partial ^{m}}{\partial \alpha ^{m}}\mathcal{L}_{\Phi }^{\alpha }(0)\right]
_{\alpha =0} \\
&=&\int (\log p_{\Phi }^{(|\varphi |)}(\varphi ))^{m}P_{\Phi }^{(|\varphi
|)}(d\varphi ).
\end{eqnarray}

\begin{remark}
\label{Rem-CorrectDE}One could patch up these particular errors by
substituting \ $f_{\Phi }(\varphi )$ \ in place of \ $p_{\Phi }^{(|\varphi
|)}(\varphi )$. \ But this would not change the fact that they are secondary
errors inherited from the inherently erroneous \ $\mathcal{G}_{\Phi
}^{\alpha }(h)$ \ and \ $\mathcal{L}_{\Phi }^{\alpha }(h)$.
\end{remark}

\end{document}